\documentclass[12pt]{article} 
\font\fff=eufm10 scaled \magstep1
\def\g{\hbox{\fff g}}
\begingroup 
\newtheorem{thm}{Theorem}[section]       
\newtheorem{lemma}[thm]{Lemma}
\newtheorem{proposition}[thm]{Proposition}
\newtheorem{theorem}[thm]{Theorem}

\newtheorem{definition}[thm]{Definition}

\newtheorem{remark}[thm]{Remark}          
\newtheorem{observation}[thm]{Observation}        
\endgroup
\def\matel#1.#2.{{\displaystyle \left\langle #1 \atop #2 \right\rangle}}
\def\state#1.{|#1\rangle}
\def\.#1.{\langle #1 \rangle_{wv}}
\def\qed{\qquad\framebox[7pt]\medskip\noindent}

\def\half{{\textstyle\frac{1}{2}}}
\newenvironment{proof}{{\sl Proof:}\quad}{\hfill{\qed}\\ \noindent}
\renewenvironment{abstract}{\begin{quote}{\bf Abstract.\
}\small}{\end{quote}\bigskip}
\title{Berezin Quantization of the Schr\"odinger Algebra}
\author{Philip Feinsilver\thanks{\leftskip -1cm Department of Mathematics
Southern Illinois University 
 Carbondale, IL. 62901, U.S.A.}, 
Jerzy Kocik\thanks{\leftskip -1cm Department of Mathematics
Southern Illinois University 
 Carbondale, IL. 62901, U.S.A.}, 
and Ren{\'e} Schott\thanks{\leftskip -1cm IECN and LORIA
 Universit\'e Henri Poincar\'e-Nancy 1, BP 239, 
54506 Vandoeuvre-l\`es-Nancy, \hbox to 18pt{}France.}} 

\date{}

\begin{document}

\maketitle

\begin{abstract}
We examine the Schr\"odinger algebra in the framework of Berezin
quantization. First, the Heisenberg-Weyl and sl(2) algebras are
studied. Then the Berezin representation of the Schr\"odinger algebra 
is computed. 
In fact, the sl(2) piece of the Schr\"odinger algebra can be 
decoupled from the Heisenberg component. This is accomplished 
using a special realization 
of the sl(2) component that is built from the Heisenberg piece as
the quadratic elements in the Heisenberg-Weyl enveloping algebra.
The structure of the Schr\"odinger algebra
is revealed in a lucid way by the form of the Berezin representation.



\end{abstract}

\vfill
\pagebreak
\tableofcontents
\pagebreak

\section{Introduction}

The Schr\"odinger algebra is a Lie algebra that has attracted 
since its introduction \cite{Ha, Ni} considerable interest 
in mathematical physics and its applications 
(see, e.g., \cite{BHP, BR, BX, BPPW, DDM}). 
\\

In \cite{FKS} we have investigated the semidirect product  
structure of the Schr\"odinger algebra and showed how 
it leads to representations in a Fock space realized 
in terms of canonical Appell systems.
This includes a classification of the representations
and construction of the Hilbert space of functions on which
certain commuting elements act as  self-adjoint operators.
Some associated evolution equations have been considered as well.
The notion of generalized coherent states is exploited extensively there.
\\

Here we shall take a somewhat different point of view and study
the Schr\"odinger algebra using the method of 
 ``Berezin quantization,'' which we understand from 
the rather broad point of view as developed from the original work of 
Berezin by Perelomov and others, see \cite{BZ74,BZ75a, BZ75b, Per}.
Again, the generalized coherent states play an essential r\^ole.
\\

Here is a description of the contents of this paper.
Section 2 presents the Schr\"odinger algebra. 
Section 3 contains the basics of our formulation of Berezin's theory. 
The Berezin representation for each of the Heisenberg-Weyl and sl(2) algebras 
is presented in \S4. 
The Berezin quantization of the Schr\"odinger algebra constitutes \S5. 
Concluding remarks and some further lines for research are given in \S6.

\section{Schr\"odinger algebra}
\label{sec:envalg}

The  ($n=1$, centrally-extended) Schr\"odinger algebra ${\cal S}$
is spanned by the following elements
\begin{eqnarray*}
M   &\qquad& \hbox{\rm mass} \\
K   &\qquad& \hbox{\rm special conformal transformation} \\
G   &\qquad& \hbox{\rm Galilei boost} \\
D   &\qquad& \hbox{\rm dilation} \\
P_x &\qquad& \hbox{\rm spatial translation} \\
P_t &\qquad& \hbox{\rm time translation} \\
\end{eqnarray*}
(see, e.g., \cite{DDM} for details)
which satisfy the following commutation relations 
given here in the form of a multiplication table
$$
\bordermatrix{
     & M & K    & G   & D     & P_x &    P_t \cr
M    & 0 & 0    & 0   & 0     &   0 &      0 \cr 
K    & 0 & 0    & 0   & -2\,K &  -G &     -D \cr 
G    & 0 & 0    & 0   & -G    &  -M &   -P_x \cr 
D    & 0 & 2\,K & G   & 0     &-P_x &-2\,P_t \cr 
P_x  & 0 & G    & M   & P_x   &   0 &      0 \cr 
P_t  & 0 & D    & P_x & 2\,P_t&   0 &      0 \cr }
$$
(thus, e.g., $[D,K]=2K$).
The Schr\"odinger algebra can be viewed as a semidirect product
$$
{\cal S} \cong {\cal H} \oplus_s sl(2)
$$
of two subalgebras:
a Heisenberg-Weyl subalgebra ${\cal H}=\hbox{\rm span}\,\{M, G, P_x\}$,
and $sl(2)=\hbox{\rm span}\,\{K, D, P_t\}$.
\\

This fundamental feature is considered in some detail in \cite{FKS}.

\section{Cartan decomposition and Berezin theory}
\label{sec:berezin}

Our approach to Berezin quantization \cite{BZ74, BZ75a, BZ75b}
is based on the exposition in \cite{Per} that uses the notion of 
generalized coherent state as a group element acting on an appropriate 
vacuum state, but, like in \cite{He}, goes beyond the 
``symmetric space paradigm.'' 
For more on the calculational tools used, the reader may 
consult \cite{FSVol1}.  
\\

Consider a Lie algebra $\g$ that admits a splitting
\begin{equation}\label{eq:CC}
       \g = \mathcal{L}\oplus\mathcal{K}\oplus\mathcal{P}
\end{equation}
where $\cal R$ and $\cal L$ are two abelian subalgebras of the 
same dimension $n$, such that they generate the whole algebra:
$\g = {\rm gen}\,\{\mathcal L, \mathcal P \}$. 

\begin{remark}\rm
An important case of such a structure is the \emph{Cartan decomposition} 
for symmetric Lie algebras with $\mathcal{L}$ and $\mathcal{P}$ satisfying 
$[\mathcal{L},\mathcal{P}]\subseteq\mathcal{K}$, 
$[\mathcal{K},\mathcal{P}]\subseteq\mathcal{P}$, and 
$[\mathcal{K},\mathcal{L}]\subseteq\mathcal{L}$. 
This precise structure, however, does not exist
for the Schr\"odinger algebra, which does not correspond to a
classical symmetric space.  
In fact there are two possibilities for a generalized
``Cartan decomposition.'' One satisfies the appropriate
commutation properties, but does not obey 
$\dim{\cal P} = \dim{\cal L}$, see \cite{FKS}. 
The other---a different variation on the standard
Cartan decomposition---will be used in the present context.
\end{remark}

Let $\mathcal{P}$, $\mathcal{L}$ and $\mathcal{K}$ have bases 
$\{R_j\}$, $\{L_j\}$, and $\{\rho_A\}_{1\le A\le m}$, respectively.
A typical element $X\in\g$ is of the form 
\begin{equation}
\label{eq:CC2}
      X= v'_jR_j  + u'_A\rho_{A}+ w'_jL_j
\end{equation}
for some $(2n+m)$-tuple $(v',u',w')$.
A group element can be obtained either by direct exponentiation of $X$,
or by composing exponentials corresponding to the factorization into subgroups
according to the decomposition of the Lie algebra.  Thus
\begin{equation}
\label{eq:grpel}
 e^X=\exp( v_jR_j) 
       \left( \prod_{A}\exp(u_{\dot A}\rho_{\dot A})\right)\exp(w_jL_j)
\end{equation}
We use the convention of summation over repeated indices, unless
the indices are dotted (the dot indicating no summation over $A$).
Clearly, the coordinates $(v,u,w)$ versus $(v',u',w')$ are mutually
dependent as they represent in (\ref{eq:grpel}) the same group element.\\

Now, we construct a representation space $\cal H$ for the enveloping 
algebra of $\g$ as a Fock space spanned by basis elements 
\begin{equation}
\label{eq:fockbasis}
        \state k_1,k_2,\ldots,k_n. = 
             R_1^{k_1}\cdots R_n^{k_n}\Omega
\end{equation}
where $\Omega$ is a vacuum state. The action of the algebra elements  
on the vacuum state is defined thus
%
%
\vskip.2in
\label{eq:lieact}                                      
(i)    $ \hfill\hat R_j   \Omega = R_j \Omega \hfill    $\par
(ii)   $\hfill \hat L_j   \Omega = 0 \hphantom{\Omega\Omega}    \hfill          $\par
(iii)  $\hfill  \hat \rho_A\Omega =\tau_A\Omega\hphantom{4}\hfill   $
\vskip.2in
where $\tau_A$ are constants.
Next, we equip $\mathcal{H}$ with a symmetric scalar product
in some number field, such that the ladder operators are mutually 
adjoint:
$$
         \hat R_i^*= \hat L_i
$$
The adjoint map for other elements is determined by the commutation rules.
We shall always consider the vacuum state normalized, 
$\langle \Omega|\Omega\rangle=1$. \\

In an important special case,  one assumes that only one element 
of $\mathcal K$, say $\rho_0$, acts on $\Omega$ as a nonzero constant $\tau$. 
Hence the group element specified by equation (\ref{eq:grpel}) acts on
$\Omega$ as follows 
\begin{equation}
\label{eq:grpaction}
        e^X\Omega = e^{\tau \,u}\,\exp(v_jR_j)\Omega 
\end{equation}
The system possesses two types of lowering and raising operators
(ladder operators): \emph{algebraic} and \emph{combinatorial}.
The \emph{algebraic lowering} and \emph{raising} operators are
defined simply by concatenation within the enveloping algebra
of $\g$ followed by acting on $\Omega$, that is
\begin{eqnarray}
\label{eq:fockaction}
 {\hat R}_j \psi &=& R_j \psi  \cr
 {\hat L}_j \psi &=& L_j \psi
\end{eqnarray}
for any linear combination $\psi$ of basis elements (\ref{eq:fockbasis}).
The ``hat" can be thus omitted without causing confusion.
The \emph{combinatorial raising operators}, ${\cal R}_j$, and  
\emph{combinatorial lowering operators}, ${\cal V}_j$, are defined to act 
on the basis (by definition) as follows
\begin{eqnarray*}
       {\cal R}_j\;\state k_1,k_2,\ldots,k_n. 
              &=& \state k_1,k_2,\ldots,k_j+1,\ldots k_n.\\
       {\cal V}_j\;\state k_1,k_2,\ldots,k_n. 
              &=& k_j\,\state k_1,k_2,\ldots,k_j-1,\ldots k_n.
\end{eqnarray*}
(``off-diagonal matrices").
\\

Next, the idea will be to express  the {\em algebraic} ladder operators,
$\hat L_j$, $\hat R_j$ (and hence the basis for $\g$), in terms of
the {\em combinatorial} ladder operators ${\cal R}_j$ and ${\cal V}_j$. 
\\

It is clear that the algebraic raising operators are represented 
directly by the $\cal R$'s, namely $\hat R_j={\cal R}_j$.
But the combinatorial lowering operators do not necessarily correspond 
to elements of $\g$. 
To find the representation we shall use the coherent states.
\\

\begin{definition}\rm
The system of coherent states $\cal C$ is the image of 
the subgroup generated by the (abelian) subalgebra $\mathcal R\subset \g$ 
in the Hilbert space $\mathcal H$ constructed above, namely
${\cal C}=\exp{\cal R} \cdot\Omega$ with the typical element
$$ 
     \state v. =\exp({v_jR_j})\Omega
$$
The coherent states form a manifold $\cal C$ 
parametrized by the elements of $\mathcal R$, or, equivalently, 
by coordinates $v=(v_1,\ldots, v_n)$.  
\end{definition}

\begin{observation}
\label{rem:hatrep}\rm
When restricted to coherent states, $\mathcal R_{j}$ acts as differentiation 
$\partial/\partial v_j$, while ${\cal V}_j$ acts as multiplication by $v_j$: 
\begin{eqnarray*}
{\cal R}_{j} &=& \partial/\partial v_j \qquad(\hbox{\rm on}\ {\cal C}) \\ 
{\cal V}_j   &=& v_j\cdot 
\end{eqnarray*} 
We shall use this property to determine the action of any operator 
defined as a (formal) operator function $f(\mathcal R, \mathcal V)$, 
with all $\mathcal V$'s to the right of any ${\mathcal R}_j$, 
by 
(1) moving all ${\cal R}$'s to the right of all ${\cal V}$'s in the
formula $f$, yielding the operator  $\check f(\mathcal V, \mathcal R)$,
and then 
(2) replacing $\mathcal{V}_j\to v_j$ and $\mathcal{R}_j\to
\partial/\partial v_j$.
Note that this is a formal Fourier transform combined with the 
\emph{Wick ordering}. 
The Berezin transform extends this by taking the inner
product with a coherent state $\state w.$. 
\end{observation}

The following notion will be used frequently.

\begin{definition}\rm
The \emph{Leibniz function} is a map ${\cal C}\times {\cal C} \to {\bf C}$ 
defined as the inner product of the coherent states:
$$ 
           \Upsilon_{wv} = \langle w|v \rangle 
$$
for any $v,w$ parametrizing $\cal C$.
\end{definition}

The Leibniz function can be explicitly calculated for a particular
Lie algebra as a scalar function symmetric with respect to $v$ and $w$.
The calculations are based on the adjoint structure:
we start with 
$$\langle w|v\rangle 
= \langle \exp({w_jR_j})\Omega \, |\,\exp({v_jR_j})\Omega \rangle
= \langle \Omega \,|\, \exp({w_jL_j}) \exp({v_jR_j})\Omega \rangle$$
and then use 
a formula for commuting a typical group element generated by $L$'s 
past a typical group element generated by $R$'s, that is, generally  
\begin{equation}
\label{eq:weyl}
   e^L\, e^R = e^r e^k e^l
\end{equation}
where $R\in{\cal R}$ and $L\in{\cal L}$ are general elements, 
while $r\in{\cal R}$ and $l\in{\cal L}$ and $k\in{\cal K}$ 
are functions of the coordinates of $R$ and $L$ and need to be 
computed in each particular case from the commutation relations. 
(See \S4 below for explicit examples.)
The relation (\ref{eq:weyl}) is called in the following the 
{\em Leibniz formula}.

\begin{definition}
The {\em coherent state representation} {\em (Berezin transform)} 
is defined for an operator $Q$ as 
$$
     \langle Q\rangle_{wv} = {\langle w| Q |v \rangle 
                            \over \langle w|v\rangle}\,.
$$ 
\end{definition}

The Berezin transforms of the
algebraic ladder operators can be expressed in terms of derivatives of 
the Leibniz function,
namely
\begin{eqnarray*}
\langle \hat R_j \rangle_{wv} 
            &=& \Upsilon^{-1}\frac{\partial}{\partial v_j}\Upsilon
                 =\frac{\partial(\log\Upsilon)}{\partial v_j} \\
\langle \hat L_j \rangle_{wv} 
            &=& \Upsilon^{-1}\frac{\partial}{\partial w_j}\Upsilon
                 =\frac{\partial(\log\Upsilon)}{\partial w_j}
\end{eqnarray*}
(using the fact that $L_j$ is adjoint to $R_j$).
The right-hand sides are functions of $v$ and $w$. Hence, 
by ``eliminating $w$'s,'' one may find 
a system of first-order partial differential equations satisfied
by $\Upsilon$, say 
$$ 
\frac{\partial\Upsilon}{\partial w_j} =
       \check f_j(v,\frac{\partial}{\partial v})\,\Upsilon
$$
for some operator functions $\check f_j$. 
We shall call this a system of {\em defining partial differential
equations.} As indicated in 
the discussion above,
it gives the answer to our question of the representation of 
$\hat L_j$, namely
$$
       \hat L_j = f_j({\cal R},{\cal V})
$$
\begin{remark}\rm
Note that the converse holds as well: if we have $\hat L_j$
expressed via ${\cal R}$ and ${\cal V}$,
then $\Upsilon$ satisfies the corresponding partial differential
equation. In some cases, this can be used to find $\Upsilon$.\\

Also, note that in the case of symmetric spaces, $\log\Upsilon$ is the 
{\it K\"ahler potential}.

\end{remark}

One goal is to identify in our representation a set of $n$ mutually commuting 
self-adjoint operators $X_j$ --- observables --- which provide physical or
probabilistic 
interpretations for certain elements of the Lie algebra.  
For instance, they generate a unitary group, 
$\exp(i\sum_js_jX_j)$, for $s=(s_1,\ldots,s_n)\in {\bf R}^n$ where
$i=\sqrt{-1}$.  The scalar function defined by
\begin{equation}\label{eq:charfunction}
      \phi(s) = \langle \Omega| \exp(i\sum_js_jX_j)|\Omega \rangle 
\end{equation}
is positive-definite, so, by {\it Bochner's Theorem}, 
$\phi(s)$ is the Fourier transform of a positive measure, which is, in 
fact, the joint spectral
density of the observables $(X_1,\ldots,X_n)$.\\
\\

\section{Berezin quantization in action}

Now we shall see how these general ideas are executed in the case 
of the Heisenberg-Weyl and sl(2) algebras.
How these results appear combined in the case of the Schr\"odinger algebra
will be shown in the following Section.

\subsection{The Heisenberg-Weyl algebra}

First we define a standard form of the Heisenberg-Weyl algebra.

\begin{definition}\rm
The standard basis for a Heisenberg-Weyl (HW) algebra
${\cal H} \cong \hbox{\rm span}\,\{P,X,H\}$ satisfies 
$$ 
     [P,X]=H, \quad [P,H]=[X,H]=0 
$$
Given a scalar $m>0$, an $m$-HW algebra denotes a representation 
where $H$ acts as the scalar $m$ times the identity operator.
\end{definition}

The Leibniz formula for the $m$-HW algebra is
\begin{displaymath}
\label{eq:HWLeib}
     e^{w P}e^{v X} = e^{v X} e^{m\,wv} e^{w P}
\end{displaymath}
(known in the literature as the Weyl formula 
and essential in quantum mechanics).
The Hilbert space (Fock space), has basis $\state n.=X^n\Omega$
with rules $P\Omega=0$, $H\Omega=m\Omega$.
>From the equation above and the relation $P=X^*$, 
the Leibniz function can be easily calculated:
$$
     \Upsilon_{wv}  = \langle e^{wX}\Omega|e^{vX}\Omega \rangle
                    = \exp(mwv)
$$
where we assume a normalized vacuum state.
Then, the following {\em defining partial differential equation}
$$ 
     \frac{\partial\Upsilon}{\partial w} = mv\Upsilon
$$
suggests how the algebra basis can be expressed in terms of 
the combinatorial  ladder operators ${\cal R}$, ${\cal V}$, namely
$$ 
     \hat X = {\cal R},\quad \hat H = m,\quad \hat P = m{\cal V}
$$

\begin{remark}\rm
Note that in this case, we could actually find the action of $P$ directly 
using the adjoint action of the group: 
\begin{eqnarray*}
P\,|v\rangle  &=&  Pe^{vX}\Omega=e^{vX}e^{-vX}Pe^{vX}\Omega \cr
              &=&e^{vX}e^{-v{\rm ad\,}X}P\Omega=e^{vX}(P+mv)\Omega \cr
              &=&mve^{vX}\Omega = m {\cal V} \,|v\rangle 
\end{eqnarray*}
\end{remark}

For the Berezin transforms, first we have
$\.X. = \Upsilon^{-1}\partial\Upsilon/{\partial v}=mw$
so, from $P=X^*$, exchanging $w\leftrightarrow v$, we immediately have 
$\.P.=mv$. Summarizing, 

\begin{proposition}
For the $m$-HW algebra, the Berezin representation is
$$  
      \.X. =mw,\quad \.H.=m,\quad  \.P.=mv
$$
\end{proposition}

Note that the Berezin representation of the operator $X_1=X+P$ 
is $\.X_1. = m(w+v)$, which, being symmetric in $w$ and $v$, is 
formally self-adjoint.\\

Recall the exponential formula
\begin{equation} \label{eq:HWexp}
    \exp(aP_x+bG)= e^{bG}\exp(mab/2)e^{aP_x}
\end{equation}
which can be readily verified by differentiation and the adjoint
action. Then we have the function $\phi(s)$, as in equation
(\ref{eq:charfunction}), 
$$
\phi(s) = \langle\Omega|\exp(isX_1)|\Omega\rangle = e^{-s^2m/2}
$$
which is the well-known Fourier transform of a normal
distribution with density function $\exp(-x^2/(2m))/\sqrt{2\pi m}$.
Thus, we interpret $X_1$ as a Gaussian random variable with variance $m$.

\subsection{The sl(2) algebra}

Now we proceed similarly with the algebra sl(2).

\begin{definition}\rm
The standard basis of the sl(2) algebra 
${\cal A} \cong \hbox{\rm span}\;\{L,R,\rho\}$
satisfies
$$ 
       [L,R]=\rho,\quad [\rho,R]=2R,\quad [L,\rho]=2L
$$
\end{definition}

In this basis, the sl(2) Leibniz formula is
\begin{equation}
\label{eq:sl2Leib}
e^{wL}e^{vR} = \exp(\frac{v}{1-wv}\,R)
(1-wv)^{-\rho} \exp(\frac{w}{1-wv}\,L)
\end{equation}
This can be computed using differential equations, as in 
\cite{FSVol1}, 
or using $2\times2$ matrices, cf. \cite{FSVol3}. 
\\

Our Hilbert space has basis $\state n.=R^n\Omega$ with  
the rules $L\Omega=0$, and $\rho\Omega=c\Omega$ for a constant $c$.
With $L=R^*$, the Leibniz function follows easily from the
Leibniz formula:
$$
     \Upsilon_{wv} = \langle e^{wR}\Omega|e^{vR}\Omega \rangle
                   = (1-wv)^{-c}
$$
The Leibniz function satisfies the partial differential equation
$$
\frac{\partial\Upsilon}{\partial w} = 
cv\Upsilon+v^2\,{\partial\Upsilon\over\partial v}
$$
from which one can read the following representation of the 
algebra in terms of the combinatorial operators ${\cal R}$, ${\cal V}$:
$$ 
    \hat R = {\cal R},\qquad 
    \hat L = c{\cal V}+{\cal R}{\cal V}^2
$$
(As in the HW case, we alternatively can find the action of $L$ via
the adjoint action of the group.)\\
To find $\hat\rho$, we calculate $[L,R]$ to get:
$$
   \hat\rho=[\hat L,\hat R]
           =[c{\cal V}+{\cal R}{\cal V}^2,{\cal R}]
           =c+2{\cal R}{\cal V}
$$
For the Berezin transforms, we have
\begin{proposition}
The Berezin representation of the sl(2) algebra is
$$
   \.R.=\frac{cw}{1-wv},\quad \.\rho.
       =c\,\frac{1+wv}{1-wv},\quad \.L.
       =\frac{cv}{1-wv}
$$  
\end{proposition}
\begin{proof}
The transform for $R$ comes directly by taking the logarithmic
derivative of $\Upsilon$ with respect to $v$. Then the result for $L$
follows as it is adjoint to $R$. For $\rho$, convert
$\hat\rho=c+2{\cal R}{\cal V}$ to 
$c+2v \partial\log\Upsilon/ \partial v$ to find the stated result.
\end{proof}

Now consider $X_2=R+\rho+L$. We have
$$ 
      \.X_2. = c\,\frac{(1+w)(1+v)}{1-wv}
$$
which is symmetric in $w$ and $v$, showing that $X_2$ defines a  
formally self-adjoint operator.
\\

\section{Berezin quantization of the Schr\"odinger algebra}
\label{sec:BerezSchr}

We now will see how the results of the previous section relate to 
the Schr\"odinger algebra $\cal S$.
\\

First we find the Leibniz formula and the Leibniz function 
for $\cal S$, and then the Berezin representations of its basis elements. 
Next we will identify two (essentially) self-adjoint operators 
acting in the Hilbert space for $\cal S$. 
Further investigation of the Berezin representation will lead
to a structure theorem for the Schr\"odinger algebra.
\\

We consider the following decomposition of the Schr\"odinger algebra
(braces represent spanning)
$$
{\cal S} = 
\underbrace{\{\,P_x, P_t\,\} }_{\cal L}
\oplus
\underbrace{\{\,M, \rho\,\} }_{\cal K}
\oplus
\underbrace{\{\,K, G\,\} }_{\cal P}
$$
Thus, in terms of the notation of Section \ref{sec:berezin}, we associate
$R_1=K$, $R_2=G$, $L_1=\,P_t$ and $L_2=P_x$.   
\\

The Hilbert space $\cal H$ is a Fock space defined as
$$
   {\cal H} = \overline{\rm span}\,
               \{\; \state n_1,n_2.\equiv K^{n_1}G^{n_2}\Omega \;\} 
$$
the bar indicating closure of the linear span, 
and the algebra elements act on the vacuum state $\Omega$ as follows
$$ 
      P_x\Omega=P_t\Omega= 0,\qquad 
        D\Omega=c\Omega, \qquad
        M\Omega = m\cdot \Omega
$$
so that on any element of $\cal{H}$, $M$ acts as multiplication by the
scalar $m$.
We assume that $\cal H$ is equipped with an inner product 
such that $K^*=P_t$ and $G^*=P_x$. The consistency of this ---
existence of such an inner product --- follows from the symmetry of
the Leibniz function calculated below.
\\

This construction makes the following two operators essentially self-adjoint:
\begin{eqnarray*}
    X_1 &=& P_t + D + K \cr
    X_2 &=& G + P_x
\end{eqnarray*}

The system of coherent states $\cal C$ is defined as a two-parameter 
manifold in $\cal H$ with typical element
$$
      \state v.=\state v_1,v_2.=e^{v_1K}e^{v_2G}\Omega
$$

\begin{lemma}\label{lem:leib}
The Leibniz formula for the Schr\"odinger algebra, reduced by
acting on the vacuum state, is 
\begin{eqnarray*}
   && e^{w_1P_t+w_2P_x}e^{v_1K+v_2G}\Omega \cr
   &&\qquad\qquad\qquad
     = (1-w_1v_1)^{-c}\,\exp(\frac{m}{2}\,\tilde q(w,v))
                        \exp(\tilde v_1 K+(\tilde v_2+w_2\tilde v_1)G)
\end{eqnarray*}
where 
\begin{eqnarray*}
\tilde v_i    &=& v_i/(1-w_1v_1), \qquad i=1,2 \\
\tilde q(w,v) &=& \frac{m}{2}\, \frac{w_1v_2^2+2w_2v_2+w_2^2v_1}{1-w_1v_1}
\end{eqnarray*}
\end{lemma}

\begin{proof}
We take several steps to pull the $P$-factors across all terms.
\begin{enumerate}
\item Apply the Leibniz formula for sl(2):
$$
     e^{w_1P_t}e^{v_1K}=e^{\tilde v_1K}(1-w_1v_1)^{-D}e^{\tilde w_1P_t}
$$
with $\tilde w_1=w_1/(1-w_1v_1)$.

\item Recall the HW formula, equation (\ref{eq:HWexp})
$$ 
    \exp(aP_x+bG)= e^{bG}\exp(mab/2)e^{aP_x}
$$

Now the adjoint action gives
$$
        e^{\tilde w_1P_t}Ge^{-\tilde w_1P_t}=G+\tilde w_1P_x
$$ 
and hence from the above HW formula,
$$
     e^{\tilde w_1P_t}e^{v_2G}\Omega
       =\exp(\frac{m}{2}\tilde w_1v_2^2)e^{v_2G}\Omega 
$$

\item Next, since $D$ acts a dilation on $G$,
$$
     (1-w_1v_1)^{-D}e^{v_2G}\Omega=(1-w_1v_1)^{-c}e^{\tilde v_2G}\Omega
$$

\item Now for the $P_x$-factor, the adjoint action gives
$$
     e^{w_2P_x}Ke^{-w_2P_x}=K+w_2G+mw_2^2/2
$$
and exponentiating,
$$
     e^{w_2P_x}e^{\tilde v_1K}
       =e^{\tilde v_1K}e^{\tilde v_1w_2G}
           \exp(m\tilde v_1w_2^2/2)e^{w_2P_x}
$$

\item And the HW Leibniz formula is the last step:
$$
        e^{w_2P_x}e^{\tilde v_2 G}
             =e^{\tilde v_2 G}\exp(mw_2\tilde v_2)e^{w_2P_x}
$$
\end{enumerate}

Combining the factors involving $m$, $K$, and $G$ yields the result
stated.
\end{proof}

This formula now yields the Leibniz function.
 
\begin{proposition}
\label{prop:leibf}
The Leibniz function for the Schr\"odinger algebra is:
\begin{displaymath}
     \Upsilon_{wv} 
       =  (1-w_1v_1)^{-c}
          \exp\Big({m\over 2}\,
          {w_1v_2^2+2w_2v_2+w_2^2v_1\over 1-w_1v_1}\Big)
\end{displaymath}
\end{proposition}
\begin{proof}
Apply the Leibniz formula in
$$
      \Upsilon_{wv}=\langle w | v\rangle 
         = \langle \Omega| e^{w_2P_x} 
           e^{w_1 P_t}\,e^{v_1K}e^{v_2G}\Omega\rangle
$$
and use the fact that appropriate elements of $\cal L$ and $\cal P$ 
are mutually adjoint, specifically, $K^*=P_t$ and $G^*=P_x$.
\end{proof}

Clearly, $\Upsilon$ is symmetric in $w$ and $v$ which shows the 
symmetry property of the inner product.\\

Now for the Berezin transforms of the Lie algebra elements, including 
the self-adjoint $X$-operators. 
The above proposition implies the following system 
of partial differential equations:
\begin{eqnarray*}
 {\partial\Upsilon\over \partial w_1}
      &=&    v_1^2\,{\partial\Upsilon\over\partial v_1}
           + v_1v_2{\partial\Upsilon\over \partial v_2}
           + cv_1\Upsilon
           + {m\over 2}\,v_2^2\Upsilon \\ 
{\partial\Upsilon\over \partial w_2}
      &=& v_1\,{\partial\Upsilon\over\partial v_2}+mv_2\Upsilon
\end{eqnarray*}
from which we can infer the hat-representation of our Lie algebra
\begin{eqnarray*}
\label{eq:hatsl2}
\hat P_t &=& c{\cal  V}_1+{\textstyle{m\over 2}}\,{\cal V}_2^2 
             + ({\cal R}_1{\cal V}_1 
             +{\cal R}_2{\cal V}_2){\cal V}_1 \cr
\hat K   &=&  {\cal R}_1 \cr
\hat D   &=&  c+2{\cal R}_1{\cal V}_1+{\cal R}_2{\cal V}_2 \cr
\hat P_x &=& m{\cal V}_2+{\cal R}_2{\cal V}_1 \cr
\hat G   &=& {\cal R}_2  \cr
\hat M   &=& m \cr
\end{eqnarray*}

To get $\hat D$, we used the commutation rule $D=[P_t,K]$.  
As a result of these calculations, the following Berezin representation 
emerges
\begin{eqnarray*}
 \.P_t. &=& c\,{v_1\over 1-w_1v_1}
            +{m\over2}\,\Big({w_2v_1+v_2\over 1-w_1v_1}\Big)^2 \\ 
 \.P_x. &=& m\,{w_2v_1+v_2\over 1-w_1v_1} \\ 
 \.K.   &=& c\,{w_1\over 1-w_1v_1}+{m\over  2}\,
            \Big({w_2+w_1v_2\over 1-w_1v_1}\Big)^2 \\ 
 \.G.   &=& m\,{w_2+w_1v_2\over 1-w_1v_1} \\ 
 \.D.   &=& c\,{1+w_1v_1\over 1-w_1v_1}
            + m\,{(w_2v_1+v_2)(w_2+w_1v_2)\over (1-w_1v_1)^2} \\
\.X_1.  &=& c\,{(1+w_1)(1+v_1)\over 1-w_1v_1}
            + m\,\Big({w_2+v_2+w_1v_2+w_2v_1\over 1-w_1v_1}\Big)^2 \\ 
\.X_2.  &=& m\,{v_2+w_2+w_1v_2+w_2v_1\over 1-w_1v_1}
\end{eqnarray*}
where the transforms for $X_i$ are found by adding the appropriate
results. In this form it is clear that indeed $K^*=P_t$, $G^*=P_x$,
$D^*=D$ and $X_i^*=X_i$, which verifies the validity of the Hilbert 
space constructed.   
The case $m=0$ recovers the sl(2) case, cf. \S4.2.
However, the $c=0$ case is interesting, as it is
unlike either of the Heisenberg-Weyl or the sl(2) cases. 
The Berezin representation of $X_2$ shows 
that it is not simply an independent Gaussian, 
which would look just like $m$ times the sum $w_2+v_2$, cf. the
operator $X_1$ in \S4.1.\\

The above formulas suggest that one should perform the following
subtractions:
$$ 
  P_t-P_x^2/(2m),\quad 
  K-G^2/(2m),\quad 
  D-GP_x/m-\half
$$
the $\half$ arises naturally as will be seen shortly.
We start with 

\begin{proposition}\label{prop:hatrep}
In the hat-representation, define 
${\cal R}_0={\cal R}_1-{\cal R}_2^2/(2m)$. Then
\begin{eqnarray*}
 \hat P_t-\hat P_x^2/(2m)  &=& (c-\half)\,{\cal V}_1+{\cal R}_0{\cal V}_1^2\\
 \hat K-\hat G^2/(2m)  &=& {\cal R}_0  \\
 \hat D-\hat G \hat P_x/m-\half &=& (c-\half)+2{\cal R}_0{\cal V}_1
\end{eqnarray*}
\end{proposition}
\begin{proof}
These follow readily from the commutation relations for the 
${\cal R}$ and ${\cal V}$ operators.
\end{proof}

Now, we find

\begin{theorem} 
\label{thm:subtractions} 
For the Schr\"odinger algebra, we have
\begin{eqnarray*}
   \.P_t-\frac{1}{2m}\,P_x^2. 
             &=& (c-\half)\,\frac{v_1}{1-w_1v_1} \\
   \.K-\frac{1}{2m}\,G^2. 
             &=& (c-\half)\,\frac{w_1}{1-w_1v_1} \\
   \.D-\frac{1}{m}\,GP_x-\half. 
             &=& (c-\half)\,\frac{1+w_1v_1}{1-w_1v_1} 
\end{eqnarray*}
Consequently, $\displaystyle L_0=P_t-\frac{1}{2m}\,P_x^2$, 
$\displaystyle R_0=K-\frac{1}{2m}\,G^2$, 
$\displaystyle \rho_0=D-\frac{1}{m}\,GP_x-\half\phantom{\biggm|}$ 
form a standard basis of an  sl(2) algebra.
\end{theorem}

\begin{proof}
Use the hat-representation from Proposition \ref{prop:hatrep} in
the dual form acting on the Leibniz function. 
Setting $L_0$, $R_0$, and $\rho_0$ as in the statement of the Theorem,
\begin{eqnarray*}
\check L_0 &=& (c-\half)\, v_1+v_1^2\,
\left(\frac{\partial}{\partial v_1}-\frac{1}{2m}\,
\frac{\partial^2}{\partial v_2^2}\right)
\\
\check  R_0 &=& \left(\frac{\partial}{\partial v_1}-\frac{1}{2m}\,
\frac{\partial^2}{\partial v_2^2}\right)  \\
\check \rho_0 &=& (c-\half)+2v_1\,
\left(\frac{\partial}{\partial v_1}-\frac{1}{2m}\,
\frac{\partial^2}{\partial v_2^2}\right)
\end{eqnarray*}
and compute accordingly.
\end{proof}

As to the main structure of the Schr\"odinger algebra, we have

\begin{theorem}
The elements $L_0,R_0,\rho_0$ defined in Theorem
\ref{thm:subtractions} commute with the Heisenberg-Weyl subalgebra
generated by $P_x,G,m$.
\end{theorem}
\begin{proof}
Use the hat-representation found in Proposition \ref{prop:hatrep}. 
With $\hat P_x=m{\cal V}_2 +{\cal R}_2{\cal V}_1$ and
$\hat G = {\cal R}_2$, it is readily checked that each of the sl(2)
operators commutes with $P_x$ and $G$.
\end{proof}

\section{Concluding remarks}

\begin{itemize}
\item For the case $n>1$, an interesting approach would be to study
representations induced from (the Lie algebra of) the Euclidean group.
On the other hand, the rotation subgroup splits off by subtracting
operators of the form $G_iP_j-G_jP_i$ from the $J_{ij}$ rotation
operators (cf. \cite{FKS}). But dealing with the representations
induced from the rotation subgroup requires some more detailed work
(cf., ``intrinsic'' subalgebras in Hecht \cite{He}).\\

\item Thanks to the decoupling structure, extending our approach 
to the $q$-Schr\"odinger algebra looks quite reasonable.\\

\item Finding the finite-dimensional representations is another project to
be considered.
\end{itemize}


\end{document}